\documentclass[english,prl,twocolumn,english,showpacs]{revtex4}
\usepackage{graphicx}
\usepackage{amsmath}
\usepackage{amssymb}
\begin{document}
\newcommand{\ahubble}{\,\frac{\dot{a}}{a}\,}
\newcommand{\overa}{\,\frac{1}{a}\,}
\newcommand{\uvec}{\boldsymbol{u}}
\newcommand{\Uvec}{\boldsymbol{v}}
\newcommand{\vvec}{\boldsymbol{v}}
\newcommand{\kvec}{\boldsymbol{k}}
\newcommand{\xvec}{\boldsymbol{x}}
\newcommand{\zvec}{\boldsymbol{z}}
\newcommand{\pvec}{\boldsymbol{p}}
\newcommand{\bfnabla}{\boldsymbol{\nabla}}
\newcommand{\bfbfield}{\boldsymbol{B}}
\newcommand{\bfefield}{\boldsymbol{E}}
\newcommand{\bfbeta}{\boldsymbol{\beta}}
\newcommand{\bfcdot}{\boldsymbol{\,\cdot\,}}
\newcommand{\bfvfield}{\boldsymbol{v}}
\newcommand{\cs}{c_{\mbox{\scriptsize{S}}}}
\newcommand{\ca}{c_{\mbox{\scriptsize{A}}}}
\newcommand{\fpi}{4\pi}
\newcommand{\upla}{\boldsymbol{u}_{\mbox{\scriptsize{P}}}}
\newcommand{\bfufield}{\boldsymbol{u}}
\newcommand{\rhop}{\rho_{\mbox{\scriptsize{P}}}}
\newcommand{\Jplas}{\boldsymbol{J}}
\newcommand{\jfield}{\boldsymbol{J}}
\newcommand{\mfi}{\boldsymbol{B}}
\newcommand{\bfield}{\boldsymbol{B}}
\newcommand{\efi}{\boldsymbol{E}}
\newcommand{\ung}{\boldsymbol{u}_{\mbox{\scriptsize{NG}}}}
\newcommand{\rhong}{\rho_{\mbox{\scriptsize{NG}}}}
\newcommand{\udm}{\boldsymbol{u}_{\mbox{\scriptsize{DM}}}}
\newcommand{\rhodm}{\rho_{\mbox{\scriptsize{DM}}}}
\newcommand{\ud}{\mathrm{d}}
\newcommand{\ApJS}{ApJ Suppl.}
\newcommand{\apjs}{ApJ Suppl.}
\newcommand{\mnras}{MNRAS}
%\newcommand{\jcph}{J.Comp.Phys.}
%\newcommand{\ARAA}{ARA\&A}
%\newcommand{\AsAs}{A\&A}
%\newcommand{\PASP}{PASP}
%\newcommand{\pasp}{PASP}

%\preprint{APS/123-QED}

%\title{How did the Magnetic Strutures collapse at the time of
%Recombination?}% Force line breaks with \\
%\title{Can we just unplug the Universe at Recombination?}
\title{Unplugging the Universe: the neglected electromagnetic consequence of
decoupling}
\author{Declan A. Diver and   Lu\'{\a i}s F. A. Teodoro }
\affiliation{Department of Physics and Astronomy, University of
Glasgow, Glasgow, G12 8QQ, Scotland, UK
}%
\email[Enquiries to ]{d.diver@physics.gla.ac.uk}
\date{\today}%

\begin{abstract}
This letter concentrates on the non-equilibrium evolution of
magnetic field structures at the onset of recombination, when the
charged particle current densities decay as neutrals are formed.

We consider the effect that a decaying magnetic flux has on the
acceleration of particles via the transient induced electric
field. Since the residual charged-particle number density is small
as a result of decoupling, we shall consider the magnetic and
electric fields essentially to be imposed, neglecting the feedback
from any minority accelerated population.

We find that the electromagnetic treatment of this phase
transition can produce energetic electrons scattered throughout
the Universe. Such particles could have a significant effect on
cosmic evolution in several ways: (i) their presence could delay
the effective end of the recombination era; (ii) they could give
rise to plasma concentrations that could enhance early
gravitational collapse of matter by opposing cosmic expansion to a
greater degree than neutral matter could; (iii) they could
continue to be accelerated, and become the seed for reionisation
at the later epoch $z \approx 10$.
\end{abstract}

\maketitle

%\section{Introduction}
%{\bf Introduction}
In this article, we consider the effect that a decaying magnetic
flux has on the acceleration of particles via the transient
induced electric field. We do not address the origin of the
magnetic field, but merely assume that it exists (see
\cite{Grasso:2001} and the references therein).

The Universe underwent recombination  at redshift $z\sim 1100$, in
which the majority of charged particles (i.e. the
current-carriers) disappeared as neutrals were formed. This can be
understood in the context of simple cosmological models (e.g.
\cite{Bean:2003PhRvD..68h3501B,Spergel:2006astro.ph..3449S,Bennett:2003ApJS..148....1B}).
In general, such models do not attempt to describe in detail the
actual non-equilibrium processes themselves, but instead simply
accommodate their broad consequences.

However there is always a risk, when presenting a series of
(quasi) equilibrium states as an evolutionary sequence, of
overlooking some essential aspect of the physics that permits the
very transition between states. In this article we develop the
consequences of describing the transient physics that occurs at
decoupling: current densities that decay must inevitably cause an
induced electric field, which must accelerate any remaining
ambient charged particles. We shall show that this physics cannot
be contained within the fluid plasma models that are
conventionally used to describe the transition; this is why such
essential physics is missing from the overall understanding of
universal evolution.

{\it Electromagnetism versus fluid models.---}  Electromagnetism
tells us that magnetic fields are sustained by displacement or
charged particle currents. Prior to recombination the high
conductivity of the plasma makes the latter the most plausible
source of magnetic curvature \cite{Ahonen:1996PhLB..382...40A,
Baym:PhysRevD.56.5254}. After decoupling only  a small fraction of
electrons $x_e \sim 10^{-5}$ \cite{Padmanabhan:1993} persists,
meaning that the continuity of the total current must inevitably
involve the displacement current replacing the particle current,
and consequently the evolution of the pre-decoupling magnetic
structures. The accompanying induced electric field will
accelerate any remnant charged particles, providing a minority
source of energetic plasma that may well be significant in the
onward evolution of the cosmos.

Suppose that a typical magnetic field $\bfbfield_0$  pervades a
region of characteristic dimension $L_0$ and that the particle
current which is sustaining it vanishes. How does the system
evolve from this point onwards? Electromagnetism tells us that if
the particle currents are interrupted, then the overall system
will react with an induced electric field in an attempt to oppose
the change. The size of the induced electric field is related to
the timescale over which the magnetic flux is changing. The
subtlety here is ensuring that cosmological models permit the
appropriate timescales in the evolutionary processes. Fully
electromagnetic descriptions are entirely appropriate, but those
that depend on magnetohydrodynamical (MHD) plasmas may well be
unable to accommodate the necessary physics at transition, and as
a result, underestimate the significance of the decoupling itself.

In the fully electromagnetic description, the electric and
magnetic fields $\bfefield$, $\bfbfield$ and plasma current
density $\Jplas$ are governed by
\begin{equation}
\bfnabla \times \bfefield = - \dot \bfbfield,\quad \bfnabla \times
\bfbfield = \mu_0 \Jplas + \dot \bfefield/c^2, \label{eq:fullmagb}
\end{equation}
where  $\dot{\,}$ denotes $\partial/\partial t$. Assuming that the
current density $\Jplas$ is negligible as a result of decoupling
leads to $\bfnabla \times \bfbfield$ being determined by $\dot
\bfefield/c^2$. The dimensional analysis of (\ref{eq:fullmagb})
%and (\ref{eq:fullmagb})
yields
%\begin{eqnarray}
%\frac{E}{L_0}  = \frac{B_0}{\tau_1}   \Rightarrow  E = \frac{L_0}
%{\tau_1} B_0, \\
%\frac{B_0}{L_0} = \frac{E}{c^2\,\tau_1} \Rightarrow
%E=\frac{c^2\tau_1} {L_0} B_0,
%\end{eqnarray}
%which implies
a timescale for changes in the field evolution of
\begin{equation}
\tau_1=L_0/c.\label{emtimescale}
\end{equation}
Hence the magnetic field reconfigures by changes which propagate
at light speed.

However, if the plasma is modelled using MHD such rapid
communication is not permitted, since the displacement current is
omitted from MHD dynamics \cite{teodoro:2007arxiv}. Hence instead
of the full electromagnetic equations, the fluid plasma of
resistivity $\eta$ is governed by
%\begin{eqnarray}
%\bfnabla \times \bfefield = - \dot \bfbfield, \quad \bfnabla \times \bfbfield &=& \mu_0 \Jplas ,   \label{eq:resmhde}\\
%\bfefield+\bfufield\times \bfbfield &=& \eta \Jplas .
%\label{eq:resmhdj}
%\end{eqnarray}
\begin{equation}
\begin{array}{cc}
\bfnabla \times \bfefield = - \dot \bfbfield , &  \bfnabla \times
\bfbfield = \mu_0 \Jplas ,   \label{eq:resmhde}
\end{array}
\end{equation}
\vspace{-0.35in}
\begin{equation}
\bfefield+\bfufield\times \bfbfield = \eta \Jplas,
\label{eq:resmhdj}
\end{equation}
in which $\bfufield$ is the plasma velocity. Note that now we
cannot ignore $\Jplas$ in favour of the displacement current,
since the former is the only way in which $\nabla \times
\bfbfield$ is sustained. It is clear at this stage that the sudden
loss of electrical current as neutrals form is beyond the compass
of such a framework; the magnetic field can only communicate
changes at the resistive or fluid-dynamic timescales. This can be
seen from the form of Ohm's Law in Eq.~(\ref{eq:resmhdj}).
Combining Eq.s~(\ref{eq:resmhde}-\ref{eq:resmhdj}) we obtain
\begin{equation}
\dot \bfbfield =  \bfnabla \times (\bfufield \times
\bfbfield)-\bar\eta \bfnabla^2 \bfbfield\
\end{equation}
in which $\bar\eta=\eta/\mu_0=\mbox{constant.}$ In magnitude terms
\begin{equation}
B_0/\tau:u_0 B_0/L_0:\bar\eta B_0/L_0^2 \quad\mbox{or}\quad 1:u_0
\tau /L_0:\bar\eta\tau/L_0^2\label{rmhdtimescales}
\end{equation}
which leads to a typical resistive time scale $\tau_2$ given by
\begin{equation}
\tau_2=L_0^2/\bar\eta
\end{equation}
assumed to be the characteristic of resistive MHD; the dynamo
timescale
\begin{equation}
\tau_3=L_0/u_0
\end{equation}
is the only possibility for an ideal (perfectly conducting) MHD
plasma.

{\it The induced electric field.---} The magnitudes of the induced
electric fields for the different scenarios are given by the
induction law, which relates the induced electromotive force to
the rate of change of magnetic flux $\Phi$:
\begin{equation}
\oint \bfefield \cdot \ud \boldsymbol{l} = -\frac{\ud \Phi}{\ud t}
= \frac{\partial}{\partial t}\int \bfbfield \cdot \ud \boldsymbol
{A}\label{eqn:induction}
\end{equation}
in which the closed-loop integral is around the path enclosing the
area $A$, in the usual way. We have used partial derivatives in
time on the right-hand side because we will consider the field at
a point in space, rather than the motional emf. If we assume a
circular area of radius $L_0$ through which the magnetic flux is
linked, then we can identify the typical size of induced electric
field by dimensional analysis:
\begin{equation}
E \sim B L_0/\tau_D
\end{equation}
where $\tau_D$ characterises the time over which the magnetic
field decays at that location, that is, the timescale for the
particle current density $\Jplas$ to disappear. If we use the
timescales associated with the electromagnetic and MHD models,
then we arrive at the following estimates of the induced electric
field:
\begin{eqnarray}
\mbox{electromagnetism} & E_1 =& c\,B_0 \\
\mbox{resistive MHD}& E_2 =& L_0/\tau_2 \,B_0\,=\,\bar\eta/L_0\,B_0 \\
\mbox{ideal MHD}& E_3 =& u_0\,B_0
\end{eqnarray}
Comparing the magnitudes of the fields by forming  $E_1:E_2:E_3$
we have
\begin{equation}
cB_0\,: \bar\eta /L_0B_0:u_0 B_0 \quad\mbox{or}\quad
1\,:\bar\eta/L_0 c\,:u_0/c
\end{equation}
Since MHD (ideal or resistive) must have $u_0 \ll c$, it is clear
from this that the induced electric field permitted by the fully
electromagnetic model is the most significant

The key issue here is that the size of the induced electric field
is determined by the rate of change of the magnetic flux. Only the
electromagnetic model can accommodate the rapidly varying current
density that results from recombination (a quantum, rather than
fluid, effect). The fluid plasma models tie the evolution of the
magnetic field explicitly and directly to the fluid motion via
$\boldsymbol{u}$. This is particularly acute in ideal MHD, where
the flux linked by a fluid element can never change, and so the
only physical process that allows flux to decay overall is cosmic
expansion. There is a subtle point here in ideal MHD: the
frozen-in flux condition means that the magnetic flux at a point
can change only at the expense of plasma motion; however, a
decrease at one point leads to an increase somewhere else as the
fluid rarifies and compresses. Hence ideal MHD cannot in fact
accommodate recombination, since the latter causes an unavoidable
overall loss of magnetic flux when the current density
(essentially) vanishes - it is not simply a case of rearrangement.

Moreover, the fluid model is further compromised by the fact that
the $\boldsymbol{u}$ that appears is the plasma fluid velocity:
the very species that is being eliminated as a result of the
recombination. Hence tying the induction field to a vanishing
fluid species cannot recover the correct physics at decoupling.
This explains why the electric field produced in such models is
deemed wholly negligible.

{\it The consequence of electric induction.---} We will now
explore the consequences of an induced electric field, considering
only the electromagnetic case. It is vital to distinguish between
timescales, however, in order to avoid confusion.

The timescale for transmission of field changes (including
boundary conditions) is the electromagnetic one, given by $\tau_1$
in Eq.~(\ref{emtimescale}); we will drop the subscript $1$ from
now on.

There is also the characteristic time for the decay of the current
density, $\tau_D$, which could be very short, since it is governed
by atomic processes: for example, by the rate coefficient for the
attachment of free electrons to positive ions. The process is
strongly density dependent, and therefore influenced by local
conditions.

Finally, we can identify a timescale $\tau_R$ over which
recombination is completed across the universe. This is taken to
be of the order of $100\,$kyr $\sim 10^{12}\,$s.
\cite{Bennett:2003ApJS..148....1B}
%(cite the latest WMAP parameters).

Hence at a location, changes in the current density due to neutral
formation (evolving over the timescale $\tau_D$) are communicated
throughout that location by electromagnetic signals, which take a
characteristic time $\tau$ to reach all parts. $\tau_R$ is the
timescale for all such locations in the universe to complete this
process.

{\it A simple example: particle acceleration.---} Dimensional
analysis of Eq.~(\ref{eqn:induction}) gives the electric field
magnitude in terms of the magnetic field and the characteristic
length and time:
\begin{equation}
E=BL_0/\tau_D
\end{equation}
In order to quantify the size of the induced electric field, we
must supply typical values for $B$, $L_0$ and $\tau_D$. A typical
magnetic field value is $B\sim 10^{-12}$T
\cite{2006Sci...311..827I,Barrow:1997PhRvL..78.3610B}. The
scale-length is more challenging: this could range from $10^{11}$m
(an astronomical unit) to $10^{16}$m (a parsec), with the latter
being more realistic in the context of structure boundaries
visible in \cite{2006Sci...311..827I,Barrow:1997PhRvL..78.3610B}.

The characteristic time for the recombination to take place at a
point is not the recombination width $\tau_R$, but the timescale
that arises self-consistently from the rate equation for the
formation of neutrals. In general terms, the rate equation
governing the conversion of one species into another via
collisions with electrons takes the form
%\begin{equation}
$\dot{n}_a=\alpha n_b n_e$
%\end{equation}
where here $n_a$ plays the role of neutral number density, $n_b$
the positive ion number density, and $n_e$ the electron number
density. The rate coefficient $\alpha$ for recombination
\cite{Wenzel:1999,bhatta:1997,Seager:2000ApJS..128..407S} is
typically $10^{-16}$m$^3$s$^{-1}$, yielding
\begin{equation}
\tau_D \sim (n \alpha )^{-1}
\end{equation}
where $n$ is a typical number density at the recombination epoch.
Taking $n \sim 10^8$m$^{-3}$ yields $\tau_D \sim 10^8$s.

Suppose we have a uniform axial magnetic field $\bfbfield(\xvec)$
within some radius $L_0$ which is supported by an edge-located
current structure which vanishes at the time of recombination. (A
helpful analogy would be a magnetic solenoid.) The magnetic field
is given by $\bfbfield(\xvec) = \hat\zvec B_0 $ and
$\nabla\times\bfbfield=0$ except where the current is carried.
When the current density is switched off, the induced electric
field is non-zero everywhere (except perhaps within the original
current carrying region) as can be seen from
Eq.~(\ref{eq:fullmagb}). We shall  assume that the induced field
propagates into the uniform magnetic field region at speed $c$,
and our analysis will be concerned with the evolution of the
system once the current collapse has been communicated across
$L_0$. This is a reasonable assumption, since we wish to follow
the energetic evolution of an inertial test particle initially at
rest when the current collapses. Of course, any candidate for
acceleration by such fields will initially be in equilibrium with
the CMBR, but this straightforward case will highlight the
essential physics.

Let us assume that $\bfbfield$ retains a predominantly axial
character during the acceleration of the test particle; there will
of course be evolving magnetic curvature, but in the spirit of
keeping the problem simple, we will set this aside. Hence the
induced electric field will remain primarily azimuthal. Therefore
\begin{equation} \rho^{-1} (\partial/\partial_\rho) \left(\rho E_\phi
\right)\approx -\dot B_0 (t) \label{eqn: etheta}
\end{equation}
where $(\rho,\phi,z)$ are standard cylindrical coordinates. The
right-hand side of Eq.~(\ref{eqn: etheta}) is independent of
$\rho$ and so we can integrate immediately to get
\begin{equation}
E_\phi(\rho,\,t)\approx -\dot B_0(t)\,\rho /2 . \label{ephi}
\end{equation}
Given these approximate fields, we can now turn to the trajectory
of an electron that is subject to them. It is appropriate to
consider the single-particle behaviour here, since we are
describing the evolution of a minority population of charged
particles that escape elimination from neutral recombination. We
know that the ionization fraction drops by 5 orders of magnitude
at recombination \cite{Padmanabhan:1993}; it is reasonable then to
consider single-particle effects without taking into account any
bulk feedback from the motion of such charges. Note that the
particle being accelerated here needn't necessarily be an original
member of the particle current ensemble that sustained the
pre-decoupling magnetic field.

The relativistic Lorentz equation for an electron moving under
imposed electric and magnetic fields is
\begin{equation}
%\frac{d\pvec }{d\tau}=-\frac{e}{\gamma}\left(\bfefield +\Uvec
\ud\pvec/\ud t=-e\left(\bfefield +\Uvec \times \bfbfield \right)
%\qquad \frac{dp_0 }{d \tau }=\frac{q}{c}\Uvec\cdot \bfefield
\end{equation}
where $\pvec = \gamma m_e \Uvec$ is the electron 3-momentum, and
$\gamma= 1/\sqrt{1-v^2/c^2}$. The reference (or lab) frame here is
the one in which the magnetic field is predominantly axial, and
the electric field azimuthal, consistent with our model
configuration.

If this particle acquires the drift speed  $\uvec_d = {\bfefield
\times \bfbfield}/{B^2} $ in the lab frame then the Lorentz force
vanishes, and the particle is free, having energy $\mathcal{E}$ in
addition to its rest energy given by
%\begin{equation}
$\mathcal{E}=(\gamma-1)m_ec^2.$
%\end{equation}
We can take $\mathcal{E}$ to be the maximum energy gain for a
particle initially at rest (or in thermal equilibrium at
decoupling) before being subjected to the induced electric field
that must accompany the collapsing current density. For $E/B
\approx 5\times 10^7$ms$^{-1}$, then $\gamma \approx 1.014$ and so
such an electron has gained approximately 1.4 \% of its rest
energy, or about 7 keV, as a direct result of the decoupling
process.

This result can be achieved in a different way. Consider a thermal
electron at decoupling, moving in a Larmor orbit in our typical
magnetic field of $1\,$pT. Using $v_\perp \approx 3 \times 10^5$
ms$^{-1}$ and $\omega_c=eB_0/m_e \approx 0.1\,$Hz, the Larmor
radius $R_{L}=v/\omega_c$ at decoupling is $R_{L0}\sim 3 \times
10^6$m. Since the induced electric field is predominantly
azimuthal, the change in azimuthal speed $\delta v$ resulting from
the azimuthal acceleration is $\delta v \approx (e/m_e) E \delta
t$, leading to an increase in Larmor radius of $\delta L_R \approx
(e/\omega_c m_e) E \delta t$. Taking the outward drift speed of
the particle as the rate of change of its Larmor radius, we arrive
at $\ud R_L / \ud t \approx E/B$, as before. Substituting for $E$
using Eq.~(\ref{ephi}), we can solve for the time evolution of the
outward drift:
\begin{equation}
R_L(t) \approx R_{L0}\left(B_0/B(t)\right)^{1/2}
\end{equation}
where $B_0$ is the magnetic field just prior to decoupling. If we
assume that the current density decays exponentially with
characteristic time $\tau_D$, then it is clear that the Larmor
radius of the particle reaches 1pc after a time $t\approx 40
\tau_D$, at which point it is travelling with a radial speed $\ud
R_L /\ud t \approx R_{L0}/(2\tau_D) \approx c/6$, as before. Note
that this outward drift speed dominates the azimuthal component.

This modelling underlines the fact that the elimination of
magnetic energy at decoupling creates an electromagnetic impulse
which accelerates ambient charged particles.

{\it Conclusions.---} In this letter we model the transient
magnetic and electric fields at the time of recombination,
distinguishing between three critical timescales: (i) $\tau$, the
time taken for fields to accommodate the evolving current density;
(ii) $\tau_D$, the time for recombination to take place in a
localised region; and (iii) $\tau_R$ the time window for
decoupling to be completed across the universe.

Dimensional analysis shows that only a full electromagnetic model
allows $\tau$ to be consistent with the appropriate physics:
recombination is a relatively rapid event driven by the relevant
rate coefficients and number densities of participating species,
and not by a global fluid (MHD) model that cannot incorporate
rapid changes in current density, and in the case of ideal MHD,
cannot allow a global change in the magnetic flux. Such MHD models
inevitably link field evolution to bulk fluid motion, thereby
imposing an unacceptabe evolution timescale.

The critical features of our model can be summarised as follows. A
minority cohort of charged particles (here taken as electrons),
initially in thermal equilibrium with the matter in the Universe,
survive the recombination process at the boundary between magnetic
domains. The rapidly changing current density caused by the
elimination of the majority population of free charged particles
(recombination) induces an electric field that accelerates any
that are left over. The maximum energy that such particles can
acquire in this process occurs when they move at the drift
velocity, corresponding to an energy of approximately 1.4\% of
their rest mass.

%The consequences of our model can be significant.
We have identified a process in which a population of energetic
particles is produced as a direct result of decoupling. Such
particles, although very much a minority fraction of the neutral
number density, are clearly not in thermal equilibrium with that
majority; this may leave a signature in the CMBR, albeit at very
low frequency. Moreover they are sufficiently energetic to cause
an element of reionisation of neutral matter in the decoupling
era, should the collisional environment be appropriate, and in so
doing, perhaps delaying the end of the recombination era.

More significantly we have described a mechanism which can seed
the Universe with high-energy charged particles from a single,
localised decoupling event. Such particles could in principle be
accelerated to even greater energies by further encounters with
induced fields at different locations, since it is clear that
$\tau_R \gg \tau_D$. Hence our model may offer a route whereby
cosmic ray particles could be created.

Finally, although in a minority, the persistence of such energetic
plasma beyond the decoupling era may accelerate the condensation
of matter into early structure formation, since long-range Coulomb
forces may be sufficient to bias gravitational self-collapse.

A full numerical simulation of the magnetic field evolution,
energetic particle production and the observational consequences
is currently underway \cite{Teodoro:2007} and will be reported
later.

DAD and LFAT acknowledge support from PPARC and the Leverhulme
Trust respectively. Thanks are also due to Tobia Carozzi for
helpful comments.
%\bibliography{mnrasmnemonic,bibliographymf}%\bibliographystyle{astron}
%\bibliography{bibliographymf}

\begin{thebibliography}{10}

\bibitem{Grasso:2001}
D~Grasso and HR~Rubinstein.
\newblock Magnetic fields in the early universe.
\newblock {\em Physics Reports-Review Section of Physics Letters}, 348(3):165
  -- 266, 2001.

\bibitem{Bean:2003PhRvD..68h3501B}
R.~{Bean}, A.~{Melchiorri}, and J.~{Silk}.
%\newblock {Recombining WMAP: Constraints on ionizing and resonance radiation at
%  recombination}.
\newblock {\em \prd}, 68(8):083501--+, October 2003.

\bibitem{Spergel:2006astro.ph..3449S}
D.~N. {Spergel}, R.~{Bean}, O.~{Dor{\'e}}, M.~R. {Nolta}, C.~L. {Bennett},
  J.~{Dunkley}, G.~{Hinshaw}, N.~{Jarosik}, E.~{Komatsu}, L.~{Page}, H.~V.
  {Peiris}, L.~{Verde}, M.~{Halpern}, R.~S. {Hill}, A.~{Kogut}, M.~{Limon},
  S.~S. {Meyer}, N.~{Odegard}, G.~S. {Tucker}, J.~L. {Weiland}, E.~{Wollack},
  and E.~L. {Wright}.
%\newblock {Wilkinson Microwave Anisotropy Probe (WMAP) Three Year Results:
%  Implications for Cosmology}.
\newblock {\em ArXiv Astrophysics e-prints}, March 2006.

\bibitem{Bennett:2003ApJS..148....1B}
C.~L. {Bennett}, M.~{Halpern}, G.~{Hinshaw}, N.~{Jarosik}, A.~{Kogut},
  M.~{Limon}, S.~S. {Meyer}, L.~{Page}, D.~N. {Spergel}, G.~S. {Tucker},
  E.~{Wollack}, E.~L. {Wright}, C.~{Barnes}, M.~R. {Greason}, R.~S. {Hill},
  E.~{Komatsu}, M.~R. {Nolta}, N.~{Odegard}, H.~V. {Peiris}, L.~{Verde}, and
  J.~L. {Weiland}.
%\newblock {First-Year Wilkinson Microwave Anisotropy Probe (WMAP) Observations:
%  Preliminary Maps and Basic Results}.
\newblock {\em \apjs}, 148:1--27, September 2003.

\bibitem{Ahonen:1996PhLB..382...40A}
J.~{Ahonen} and K.~{Enqvist}.
%\newblock {Electrical conductivity in the early universe}.
\newblock {\em Physics Letters B}, 382:40--44, 1996.

\bibitem{Baym:PhysRevD.56.5254}
G.~{Baym} and H.~{Heiselberg}.
%\newblock Electrical conductivity in the early universe.
\newblock {\em Phys. Rev. D}, 56(8):5254--5259, 1997.

\bibitem{Padmanabhan:1993}
T.~Padmanabhan.
\newblock {\em Structure formation in the Universe}.
\newblock Cambridge University Press, Cambridge, U.K., 1st edition, 1993.

\bibitem{teodoro:2007arxiv}
L.F.A. Teodoro, D.A. Diver, and M.A. Hendry.
\newblock {\em A cautionary note on cosmological magnetic fields.}
\newblock astro-ph/0702729, submitted to {\em Phys. Rev. D}

\bibitem{2006Sci...311..827I}
K.~{Ichiki}, K.~{Takahashi}, H.~{Ohno}, H.~{Hanayama}, and N.~{Sugiyama}.
%\newblock {Cosmological Magnetic Field: A Fossil of Density Perturbations in
%  the Early Universe}.
\newblock {\em Science}, 311:827--829, February 2006.

\bibitem{Barrow:1997PhRvL..78.3610B}
J.~D. {Barrow}, P.~G. {Ferreira}, and J.~{Silk}.
%\newblock {Constraints on a Primordial Magnetic Field}.
\newblock {\em Physical Review Letters}, 78:3610--3613, May 1997.

\bibitem{Wenzel:1999}
U.~Wenzel, K.~Behringer, A.~Carlson, J.~Gafert, B.~Napiontek, and A.~Thoma.
%\newblock Volume recombination in divertor i of asdex upgrade.
\newblock {\em Nuclear Fusion}, 39(7):873--882, 1999.

\bibitem{bhatta:1997}
S.~Bhattacharyya, A.~Roy, and S.~Mitra.
%\newblock Radiative recombination of cold electron with proton and deuteron.
\newblock {\em Fizika A}, 6(4):149--160, 1997.

\bibitem{Seager:2000ApJS..128..407S}
S.~{Seager}, D.~D. {Sasselov}, and D.~{Scott}.
%\newblock {How Exactly Did the Universe Become Neutral?}
\newblock {\em \apjs}, 128:407--430, June 2000.

\bibitem{Teodoro:2007}
L.F.A. Teodoro and D.A. Diver.
\newblock {\em The collapse of the cosmological magnetic field at recombination.}
\newblock (in preparation)

\end{thebibliography}

\pagestyle{empty}\end{document}